\documentclass[journal=jctcce,manuscript=article,layout=onecolumn]{achemso}
\setkeys{acs}{articletitle = true}
\setkeys{acs}{chaptertitle = true}
\setkeys{acs}{abbreviations = false}
\setkeys{acs}{maxauthors=150}
\setkeys{acs}{etalmode=truncate}

\usepackage[version=3]{mhchem} 
\usepackage{subcaption}
\usepackage{xcolor}
\usepackage{siunitx}
\usepackage{hyperref}

\newcommand{\bld}{\boldsymbol}
\newcommand{\mrm}{\mathrm}
\newcommand{\ket}[1]{\vert #1 \rangle}

\newcommand{\bra}[1]{\langle #1 \vert}

\newcommand{\Tbraket}[3]{\langle #1 \hspace{.10em} \vert \hspace{.10em} #2 \hspace{.10em} \vert \hspace{.10em} #3 \rangle}

\newcommand{\SM}{\boldsymbol{S}}
\newcommand{\DM}{\boldsymbol{D}}
\newcommand{\CM}{\boldsymbol{C}}

\newcommand{\etal}{\textit{et al.}}

\author{Sarai Dery Folkestad}
\altaffiliation{These authors contributed equally to this work.}
\author{Alexander C. Paul}
\altaffiliation{These authors contributed equally to this work.}
\author{Regina Paul (née Matveeva)}
\affiliation{Department of Chemistry, Norwegian University of Science and Technology, NTNU, 7491 Trondheim, Norway}
\altaffiliation{These authors contributed equally to this work.}
\author{Peter Reinholdt}
\affiliation{Department of Physics, Chemistry and Pharmacy, University of Southern Denmark, SDU,
Campusvej 55, 5230 Odense,
Denmark}
\author{Sonia Coriani}
\affiliation{Department of Chemistry, Technical University of Denmark, DTU, Kemitorvet Bldg 207,
2800 Kongens Lyngby, Denmark}
\author{Michael Odelius}
\affiliation{Department of Physics, Stockholm University, 10691 Stockholm, Sweden}
\author{Henrik Koch}
\affiliation{Department of Chemistry, Norwegian University of Science and Technology, NTNU, 7491 Trondheim, Norway}
\email{henrik.koch@ntnu.no}

\title{X-ray Absorption Spectra for Aqueous Ammonia and Ammonium: Quantum Mechanical versus Molecular Mechanical Embedding Schemes}

\abbreviations{IR,NMR,UV}

\begin{document}

\begin{abstract}
The X-ray absorption (XA) spectra
of aqueous ammonia and ammonium are computed
using a combination of coupled cluster singles and doubles (CCSD)
with different quantum mechanical and molecular mechanical embedding schemes.
Specifically, we compare frozen Hartree--Fock (HF) density embedding, polarizable embedding (PE),
and polarizable density embedding (PDE).
Integrating CCSD with frozen HF density embedding is possible within
the CC-in-HF framework, which circumvents the conventional
system-size limitations of standard coupled cluster methods.
We reveal similarities between PDE and frozen HF density descriptions,
while PE spectra differ significantly.
By including approximate triple excitations, we also investigate the effect
of improving the electronic structure theory.
The spectra computed using this approach show an improved intensity ratio
compared to CCSD-in-HF.
Charge transfer analysis of the excitations
shows the local character of the pre-edge and main-edge,
while the post-edge is formed by excitations delocalized over the first solvation shell and beyond.
\end{abstract}

\section{Introduction}
Over the last fifty years, ammonia (NH$_3$) and ammonium (NH$_4^+$) have raised the interest of the scientific community for various reasons:
One reason is the controversy surrounding their solvation structure.\citep{ekimova2017aqueous}
Another point of interest is the rapid rotation of ammonia in liquid water,
which has been under debate for several years.\citep{perrin1986rotation}
Ammonia and ammonium are also known in the context of wastewater treatment as the
primary forms of inorganic nitrogen.~\cite{ronan2021recent}
Moreover, ammonia solution has emerged as a potentially effective CO$_2$ absorbent,
\citep{chen2012studies, feng2021adsorption}
an important feature in times of increasing environmental pollution.

Ammonia and ammonium have been the subject of numerous experimental and theoretical studies.
\citep{pullman1974ab, rode1988simulation, bruge1999ab, bacelo2002theoretical, weinhardt2011nuclear, ekimova2017aqueous, guo2020hydration, reinholdt2021nitrogen, carter2022choice, patkar2022tug}
Among these are investigations
using X-ray absorption (XA) spectroscopy.\cite{ekimova2017aqueous, reinholdt2021nitrogen}
When simulating XA spectra of liquids and solutions,
it is crucial to account for solvent effects,
as the electronic spectra of a molecule in solution are directly influenced by a solvent.
When calculating such spectra at a fully quantum mechanical level,
including a large number of solvent molecules poses a challenge.
Therefore, hybrid methods,
which involve treating solute and solvent at different levels of theory,
have been actively used to describe aqueous solutions.
These methods combine high-level quantum mechanical (QM) calculations
with more approximate models
such as polarizable continuum (PCM),
molecular mechanics (MM),
and Hartree--Fock (HF) or Kohn-Sham density embedding.\cite{sanchez2010cholesky, manby2012simple}

Recently,
Reinholdt \textit{et al.}\cite{reinholdt2021nitrogen} presented XA spectra of
NH$_3$ and NH$_4^+$ in water computed using the combination of
coupled cluster (CC) methods with polarizable embedding (PE).
Coupled cluster singles and approximate doubles (CC2) and
coupled cluster singles and doubles (CCSD) were the methods used.
The spectra obtained using CCSD showed satisfactory results
and a notable improvement compared to conventionally used
transition-potential density functional theory (TP-DFT).\cite{triguero1998calculations}
However, as the authors themselves pointed out,
it would be beneficial to employ multilevel coupled cluster methods and/or
extend the number of water molecules in the QM region to validate their conclusions.

In this work, we model the XA spectra of aqueous NH$_3$ and NH$_4^+$
using polarizable density embedding (PDE)-CCSD and
coupled cluster methods embedded in a frozen Hartree--Fock (HF) density
(CC-in-HF),\cite{folkestad2021multilevel, eTprog, folkestad2023enhanced}
and compare our results to the PE-CCSD spectra from
Ref.~\citenum{reinholdt2021nitrogen}.
The CC-in-HF calculations were performed with
a development version of the $e^\mathcal{T}$ program,\cite{eTprog} while the PDE-CCSD calculations were carried out with the Dalton program\cite{daltonpaper} (version 2022).

\section{Theory}
The coupled cluster wave function is obtained through the exponential ansatz,
\begin{align}
   \ket{ \mathrm{CC} } = e^T\ket{\mathrm{R}}~,
\end{align}
where the exponential of the cluster operator, $T$, acts on a reference determinant (typically the Hartree--Fock state). The cluster operator generates excitations of the reference, and can be written in terms of excitation operators, $\tau_\mu$, and corresponding amplitudes, $t_{\mu}$:
\begin{align}
    T = \sum_{\mu}t_{\mu}\tau_\mu = T_1 + T_2 + T_3 + \ldots + T_{N_e}~.
\end{align}
For instance, single, double, and triple excitations of the reference are generated by $T_1$, $T_2$, and $T_3$, respectively.

The coupled cluster energy is obtained by
projecting the Schrödinger equation onto the reference determinant,
\begin{align}
    \Tbraket{\mathrm{R}}{e^{-T}He^T}{\mathrm{R}} = E_\textrm{CC}~,
\end{align}
and the cluster amplitudes are
determined by projection onto the space of excited determinants {$\bra{\mu} = \bra{\textrm{R}}\tau_{\mu}^\dagger$}~,
\begin{align}    \Omega_{\mu} = \Tbraket{\mu}{e^{-T}He^T}{\mathrm{R}} = 0~. \label{eq:omega}
\end{align}

{In practice, the cluster operator must be truncated at a certain excitation level,
which defines the hierarchy of standard coupled cluster models.
Among the most commonly used models is the coupled cluster singles and doubles (CCSD) model,
 where $T = T_1 + T_2$.}
In the spin-adapted closed-shell formulation of the theory, we have
\begin{align}
    T_1 = \sum_{ai}t^{a}_i E_{ai},\quad T_2 = \frac{1}{2}\sum_{aibj}t^{ab}_{ij}E_{ai}E_{bj},
\end{align}
where $E_{ai}$ is a singlet excitation operator, $t^{a}_i$ and $t^{ab}_{ij}$ are the cluster amplitudes. We use indices $\{i, j,\ldots\}$ for occupied orbitals, $\{a, b,\ldots\}$ for virtual orbitals, and $\{p, q,\ldots\}$ to denote general orbitals.

Excitation energies and transition moments
can be calculated with linear response theory or the equation-of-motion\citep{sekino1984linear,geertsen1989equation,bartlett2012coupled} (EOM) framework. In EOM coupled cluster theory, the ground and excited states are expressed as the eigenvectors of the similarity-transformed Hamiltonian,
\begin{align}
    \bld{\bar{H}} =
    \begin{pmatrix}
    0 & \bld{\eta}^T \\
    0 & \bld{A} \\
    \end{pmatrix} + E_\mrm{CC}\bld{I},
\end{align}
where $A_{\mu\nu}=\Tbraket{\mu}{[\bar{H},\tau_{\nu}]}{\mrm{HF}}$ is the Jacobian matrix and
$\eta_{\nu} = \Tbraket{\mrm{HF}}{[\bar{H}, \tau_{\nu}]}{\mrm{HF}}$.
The ground state amplitudes have been determined from eq \eqref{eq:omega}.
The $k$'th state is determined by
\begin{align}
     \bld{\bar{H}}\boldsymbol{R}^k &= (E_\mrm{CC} + \omega_k)\boldsymbol{R}^k\\
     \bld{\bar{H}}^T\boldsymbol{L}^k &= (E_\mrm{CC} + \omega_k)\boldsymbol{L}^k,
\end{align}
where $\omega_k$ is the corresponding excitation energy (for $k>0$).
Since $\bld{\bar{H}}$ is non-Hermitian, the left, $\boldsymbol{L}^k$,
and right, $\boldsymbol{R}^k$, eigenvectors differ, and so do the state vectors,
\begin{align}
    \ket{k}=\sum_{\mu\geq0}e^{T}R^k_{\mu}\ket{\mu},\quad
   \bra{k}=\sum_{\mu\geq0}L^k_{\mu}\bra{\mu}e^{-T} .
\end{align}
We need both left and right state vectors
to calculate, e.g., oscillator strengths:\citep{stanton1993equation,koch1994calculation}
\begin{align}
    f_k = \frac{2}{3}\omega_k\Tbraket{0}{\bld\mu}{k}\cdot\Tbraket{k}{\bld\mu}{0},\label{eq:os1}
\end{align}
where $0$ denotes the coupled cluster ground states.

Core excitations often have double excitation character, and accurate modeling relies on the inclusion of triple excitations in $T$.
While CCSD scales as $\mathcal{O}(N^6)$ with system size, the exact inclusion of triple excitations, as in CCSDT,
will yield $\mathcal{O}(N^8)$-scaling equations, a cost that is prohibitive for most molecular systems.
Triple excitations are, therefore, often included perturbatively at an $\mathcal{O}(N^7)$-cost,
as in the CC3\citep{Koch1997, paul2020new} model.

In this work, we calculate the XA spectra of aqueous ammonia and ammonium.
Accurately simulating
XA spectra of solutions or liquids requires many
representative geometries and a high-level electronic structure description
of both solute and neighboring solvent molecules.
Hence, a large number of calculations on a sizable molecular system must be carried out, rendering an $\mathcal{O}(N^7)$-scaling model such as CC3 unfeasible.

\subsection{Multilevel coupled cluster theory}
Multilevel coupled cluster (MLCC) theory\cite{myhre2014multi} can be used to reduce the cost of coupled cluster calculations while retaining high accuracy in intensive properties.
In MLCC, the highest-order excitations in the cluster operator are restricted to an active space. For instance, in MLCCSDT, $T_3$ is confined to an active orbital space, whereas $T_1$ and $T_2$ include single and double excitations between
all occupied and virtual orbitals in the molecular system.
The ground and excited states are solved analogously to the standard coupled cluster models.
However, the projection space is constrained
to correspond to the excitations included in $T$.
In this paper, we use MLCC3,\cite{myhre2016multilevel, paul2022oscillator}
a multilevel analog of CC3 with triple excitations restricted to an active space.

A similar active space strategy can also define a Hartree--Fock density embedding scheme:
by restricting all cluster amplitudes to the active orbital space, we get the CC-in-HF\cite{sanchez2010cholesky,folkestad2020equation, folkestad2021multilevel}
model. For example, in the CCSD-in-HF model, both $T_1$ and $T_2$ are restricted. Since
high-scaling equations are limited to the active orbital space, we can use CC-in-HF on large molecular systems.
Unlike QM/MM approaches, the CC-in-HF models offer a single wave function for the whole molecular system under consideration.
Thereby, we include Pauli-repulsion between the target (active region)
and environment (rest of the system) regions.
In the CC-in-HF scheme, an effective Fock matrix is used to include the interactions with the environment in the coupled cluster calculation:
\begin{align}
    f^{\text{eff}}_{\mu\nu} = h_{\mu\nu} +\sum_{\gamma\delta} \Big((\mu\nu|\gamma\delta) - \frac{1}{2}(\mu\delta|\gamma \nu)\Big)D^\mathrm{target}_{\gamma\delta}
    +\sum_{\gamma\delta} \Big((\mu\nu|\gamma\delta) - \frac{1}{2}(\mu\delta|\gamma \nu)\Big)D^\mathrm{env}_{\gamma\delta}.\label{eq:Fock-CC-in-HF}
\end{align}
Here, $h_{\mu\nu}$ are the one-electron integrals in the atomic orbital (AO) basis, $(\mu\nu|\gamma\delta)$ are the electron repulsion integrals in Mulliken ordering, and $D^\mathrm{target}_{\gamma\delta}$ and $D^\mathrm{env}_{\gamma\delta}$ are the active and environment densities, respectively.
The total density,
\begin{align}
    {D}_{\mu\nu} = {D}^{\mathrm{target}}_{\mu\nu} + {D}^{\mathrm{env}}_{\mu\nu},
\end{align}
is optimized at the Hartree--Fock level of theory, and the partitioning of the orbital space occurs between the Hartree--Fock and coupled cluster calculations.
Note that all AOs---both of the target region and the environment---can contribute to the active molecular orbitals, i.e., the AO indices $\mu$ and $\nu$ are not restricted to the target region in eq \eqref{eq:Fock-CC-in-HF}. Conversely, AOs centered on the active region can contribute to the environment density. A reduced dimension of the effective Fock matrix is only obtained through transformation to the active MO basis.

Finally, we can combine the MLCC and the CC-in-HF approaches to obtain the so-called MLCC-in-HF model.
This can be done by introducing two concentric active spaces,
where all $t$-amplitudes are restricted to the outer space,
and the high-order $t$-amplitudes are limited to the inner space.\cite{folkestad2021multilevel}

Successful application of the CC-in-HF and MLCC approaches
depends on the partitioning of the orbital space.
We need to choose a suitable type of Hartree--Fock orbitals
and determine the size of the active space.
For the
XA spectra of ammonia and ammonium, we select the outer active orbital layer using localized orbitals.
This ensures that the solute and a fixed number of solvent molecules are included
in the correlated part of the calculation. Any kind of localized Hartree--Fock orbitals can be used. In this work, we use Cholesky occupied orbitals\cite{sanchez2010cholesky} and orthonormalized projected atomic orbitals\citep{pulay1983,saebo1993} (PAOs) for the virtual space. The Cholesky orbitals result
from a partial Cholesky factorization of the Hartree--Fock density in the atomic orbital basis:
\begin{align}
    {D}_{\mu\nu} = 2\sum_{i}C^i_{\mu}C^i_{\nu} + {D}^{\mathrm{env}}_{\mu\nu}.
\end{align}
The elements of the Cholesky vector, $\bld{C}^i$, are the orbital coefficients of the active occupied orbital $i$. The partial factorization is obtained by restricting the pivots of the decomposition to the atomic orbitals
of the target region. Furthermore, to ensure a tight active space, we exclude pivoting elements corresponding to diffuse orbitals with exponents smaller or equal to 0.6.

For the virtual space, active orbitals are constructed by projecting the occupied orbitals out of the atomic orbitals centered on the target region,
\begin{align}
\bld C^{\text{PAO}} = \bld I - \frac{1}{2}\bld{DS}^{\mathrm{target}}. \label{eq:pao}
\end{align}
The overlap matrix $\bld S^{\mathrm{target}}$ is rectangular and contains the columns of the AO overlap matrix that correspond to AOs centered on the active atoms.
Before the coupled cluster calculation,
we remove linear dependencies and orthonormalize the remaining virtual orbitals.

For the inner active space,
used in the MLCC3 calculations, we use correlated natural transition orbitals\citep{hoyvik2017correlated} (CNTOs). The CNTOs are constructed from CCSD (or CCSD-in-HF) excitation vectors, $\bld{R}^k$, by diagonalizing the unit-trace matrices
\begin{align}
  M^k_{ij}  = &\sum_{a} R^k_{ai}  R^k_{aj}\label{M_doubles}
  + \frac{1}{2} \sum_{abk} (1 + \delta_{ai,bk}\delta_{ij})R^k_{aibk}R^k_{ajbk}
\end{align}
and
\begin{align}
  N^k_{ab} = &\sum_{i}  R^k_{ai}  R^k_{bi}\label{N_doubles}
 + \frac{1}{2} \sum_{ijc} (1 + \delta_{ai,cj}\delta_{ab})R^k_{aicj}R^k_{bicj},
\end{align}
and transforming the occupied and virtual blocks of the orbital coefficient matrix with the resulting eigenvector matrices.
Using CNTOs, a compact representation of the excitation captured by $\bld{R}^k$ can be achieved.
To obtain an orbital space that targets a collection of excited states, the matrices
$\boldsymbol{M} = \sum_{k}\boldsymbol{M}^k$ and $\boldsymbol{N} = \sum_{k}\boldsymbol{N}^k$
are diagonalized.
The resulting eigenvector matrices are used as transformation matrices for the occupied and virtual orbital coefficients, respectively.

\subsection{Polarizable density embedding}
The polarizable density embedding\cite{olsen2015polarizable} scheme shares similarities with the Hartree--Fock density embedding outlined above, and is an extension of polarizable embedding (PE) approach by Olsen \etal\cite{olsen2010excited}
In PDE, contributions from the environment are added to the Fock matrix of the target system
according to
\begin{equation}
    \bld{f}^{\text{eff}} = \bld{f} + \bld{v}^{\text{env},\text{PDE}}.
\end{equation}
The embedding operator contains three terms, accounting for the electrostatic (es), non-electrostatic repulsion (rep), and induction (ind) interactions
\begin{equation}
    \bld{v}^{\text{env},\text{PDE}} = \bld{v}^{\text{es}} + \bld{v}^{\text{rep}} + \bld{v}^{\text{ind}}.
    \label{eq:PDE}
\end{equation}
The electrostatic term includes the effects of the nuclear-electron attraction and the
electron-electron repulsion between electrons in the target system and the nuclei and electrons of the
environment; the latter of which is contained through frozen fragment densities, $\bld{D}^f$, of the
environment molecules:
\begin{equation}
    v^\mathrm{es}_{\mu\nu} = -\sum_f\sum_{m\in f} Z_mv_{\mu\nu}(\mathbf{R}_m) + \sum_f\sum_{\gamma\delta \in f}D^f_{\gamma\delta}(\mu\nu|\gamma\delta).
    \label{eq:vnuc}
\end{equation}
Here,
$Z_m$ is the charge of nucleus $m$ in fragment $f$,
$v_{\mu\nu}$ is a nuclear attraction integral, $D_{\gamma\delta}^{f}$ is the density matrix of environment fragment $f$, and $(\mu\nu|\gamma\delta)$ are (intermolecular) electron repulsion integrals.

Similarly to the CC-in-HF approach, the PDE model retains the ``exact'' electron-electron repulsion from the electronic densities of the environment
rather than approximating this interaction with a distributed multipole model, as is done in the simpler PE models.
However, unlike in CC-in-HF, the environment density in PDE is made up of frozen fragment densities that are calculated in isolation\cite{olsen2015polarizable}; in CC-in-HF the total density ($\bld{D} = \boldsymbol{D}^{\mathrm{target}} + \boldsymbol{D}^{\mathrm{env}}$) is optimized at the Hartree--Fock level of theory. As a result, the total Hartree--Fock density ($\bld\rho = \frac{1}{2}\boldsymbol{D}$) is idempotent in CC-in-HF, which is not the case for PDE.

Compared to PE, the PDE potential also includes Pauli-repulsion through a Huzinaga-Cantu-style\cite{Huzinaga1971} repulsion operator
\begin{equation}\label{eq:rep}
     v^\mathrm{rep}_{\mu\nu} = -  \sum_f\sum_{\gamma\delta \in f} \sum_i^{N_\mathrm{occ},f}\varepsilon_i^fC^f_{\gamma i}C^f_{\delta i} S_{\mu\gamma}S_{\nu\delta},
\end{equation}
which penalizes the wave function overlap between the
target region and the environment fragment wave functions. The repulsion operator depends on the environment fragment orbital energies $\varepsilon_i^{f}$,
MO coefficients $C_{\gamma i}^f$,
and the intermolecular overlap integrals $S_{\mu\gamma}$.
This can be compared to the exchange contribution to the target system Fock matrix in CC-in-HF (see eq \eqref{eq:Fock-CC-in-HF}).

Finally, the PDE models include mutual polarization effects between the target
region and the environment
which is described by an induced dipole model, as in PE\cite{olsen2010excited}.
The induction operator appears as
\begin{equation}
    v^\mathrm{ind}_{\mu\nu}= - \sum_{m=1}^{N} \sum_{\alpha=x,y,z} {\mu}^{\mathrm{ind}}_{m,\alpha} {t}_{\mu\nu,\alpha}^{(1)} ({\mathbf{R}}_{m}). \label{eq:induced_dipole_operator}
\end{equation}
where ${\mu}^{\mathrm{ind}}_{m,\alpha}$ is a component of an induced dipole moment and ${t}_{\mu\nu,\alpha}^{(1)}$ are matrix elements of the electric field operator from the QM region. The induced dipoles are obtained as solutions to the linear problem
\begin{equation}
\boldsymbol{\mathbf{\mu}}_{m}^{\mathrm{ind}}\left(\mathbf{F}_{\mathrm{tot}}\right)=\boldsymbol{\alpha}_{m}\mathbf{F}_{\mathrm{tot}}(\mathbf{R}_{m})=\boldsymbol{\alpha}_{m}\left(\mathbf{F}(\mathbf{R}_{m})+\sum_{m'\neq s}\mathbf{T}_{mm'}^{(2)}\boldsymbol{\mathbf{\mu}}_{m'}^{\mathrm{ind}}\right),
\end{equation}
where $\mathbf{F}$ is the electric field, and $\mathbf{T}^{(2)}$ is the dipole-dipole interaction tensor.
Thus, the polarizability $\boldsymbol{\alpha}_{m}$ creates an induced dipole
in response to the electric fields from the QM density, the static fields from all other environment fragments,
and the fields from every other induced dipole moment in the environment.
As a result, the induced dipoles depend on the QM density,
which in turn depends on the values of the induced dipoles,
ensuring mutual polarization between the environment and the QM density.

\subsection{Charge transfer analysis of excitations}
When a molecular system absorbs a photon, the electrons rearrange themselves, causing a change in density,
\begin{align}
   \Delta\DM_k = \DM_k - \DM_0,
\end{align}
where $\DM_k$ is the density of an excited state and $\DM_0$ the ground state density.
The trace of the one-electron density equals the number of electrons. Therefore, the trace of
{the density difference,} $\Delta\DM_k${,} must be zero.
Using, for example, L{\"o}wdin population analysis,
we can transform the density difference
from the delocalized molecular orbital (MO) basis to the local atomic orbital (AO) basis,
\begin{align}
    \Delta\bld{\rho}^{\mrm{AO}}_k = \SM^{\frac{1}{2}} \, \CM \, \Delta\DM_k \, \CM^T \, \SM^{\frac{1}{2}},
    \label{eq:Loewdin-MO-AO}
\end{align}
where $\SM$ is the AO overlap matrix and $\CM$ contains the MO-coefficients.
Since $\CM^T \SM \CM = \bld{I}$, the trace of the density difference remains zero in the AO basis.
We can now partition the trace of $\Delta\bld{\rho}^{\mrm{AO}}_k$ into contributions from the AOs on different fragments of the molecular system. For instance, for the calculation of XA spectra of a solvent or liquid, we can partition the trace with respect to (A) AOs on the the solute, (B) AOs on the water molecules closest to the solute, and (C) the rest of the AOs:
\begin{align}
\begin{split}
   \mathrm{Tr}(\Delta\bld{\rho}^{\mrm{AO}}_k) &= \sum_{\alpha}[\Delta\bld{\rho}^{\mrm{AO}}_k]_{\alpha\alpha}\\
    &=\sum_{\alpha\in A} |\bld{\rho}\DM^{\mrm{AO}}_k]_{\alpha\alpha}
    + \sum_{\alpha\in B} |\bld{\rho}\DM^{\mrm{AO}}_k]_{\alpha\alpha}
    + \sum_{\alpha\in C} |\bld{\rho}\DM^{\mrm{AO}}_k]_{\alpha\alpha} \\
    &= 0.
\end{split}
\end{align}
The trace of a single subsystem is then interpreted
as the number of electrons detached from or attached to this subsystem,
depending on the sign.
Such analysis has been reported before
for TDDFT using Mulliken population analysis
in Ref.~\citenum{ChargeTransferNumbers0} and
more generally in Ref.~\citenum{ChargeTransferNumbers1}.

\section{Results and discussion}
Using PDE-CCSD and CC-in-HF,
we describe the solute and four closest water molecules with coupled cluster theory,
whereas the remaining solvent molecules are described
at a lower level of theory.
These methods serve to reduce the high cost associated with
applying coupled cluster methods to large systems
without compromising the accuracy in intensive molecular properties.
The calculations are performed on the 195 spherical cluster geometries
studied in Ref.~\citenum{reinholdt2021nitrogen}.
For the PDE\cite{olsen2015polarizable},
electron densities at the CAMB3LYP\cite{yanai2004new}/aug-cc-pVTZ level
are computed for each molecule except the solute and four closest water molecules.
The embedding potential is constructed from these fragment densities and
atomic polarizabilities,
obtained with the LoProp\cite{gagliardi2004local} method.
The spectra are calculated from a CCSD wave function optimized in the presence of the embedding potential.
For CC-in-HF, the orbitals localized on the solute and the four closest
water molecules are included in the coupled cluster region.
The remaining orbitals are described using a frozen HF density.\citep{sanchez2010cholesky, eTprog, folkestad2021multilevel}
We use two versions of CC-in-HF:
CCSD-in-HF and
MLCC3-in-HF (multilevel coupled cluster singles and doubles with
perturbative triples),\citep{myhre2016multilevel, paul2022oscillator}
which are described in more detail in Ref.~\citenum{folkestad2023enhanced}.
In all calculations
the core-valence separation approximation~\cite{cederbaum1987cvs,coriani2015cvs,coriani2016erratum}
is used to target core excitations.
\begin{figure*}[htb]
    \centering
    \includegraphics[width=\textwidth]{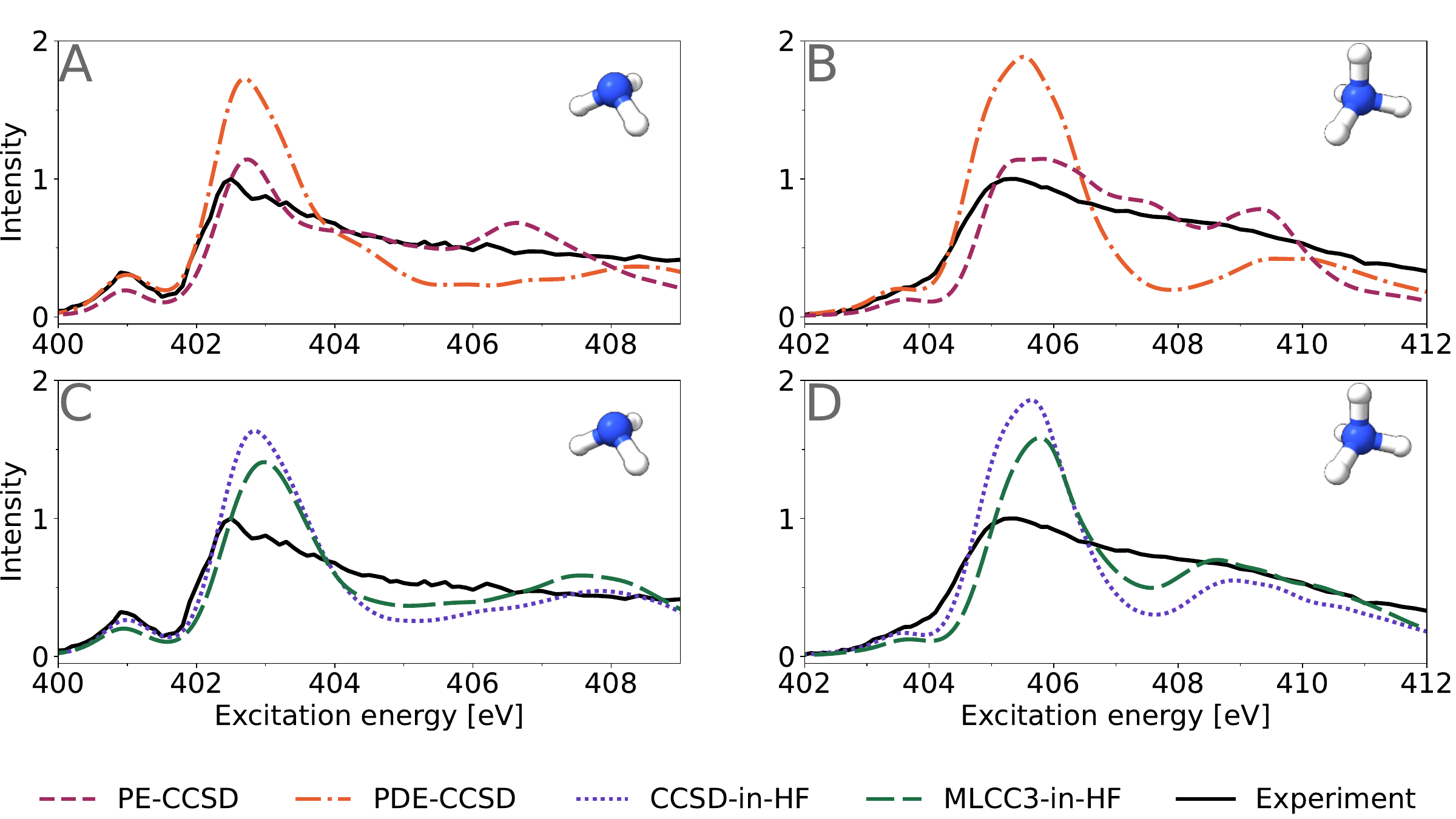}
    \caption{
        XA spectra of ammonia (panels A, C) and ammonium (panels B, D)
        in water clusters for different embedding methods.
        Solute and four closest water molecules are described with 6-311++G$^{\ast\ast}$ basis,
        while remaining solvent molecules are treated at the PE, PDE, and HF/6-31G$^{\ast\ast}$ levels of theory.
        The PE-CCSD, PDE-CCSD, CCSD-in-HF, and MLCC3-in-HF results have been shifted by
        $\SI{-1.9}{\eV}$, $\SI{-2.8}{\eV}$, $\SI{-2.5}{\eV}$, $\SI{-1.5}{\eV}$, respectively.
        The PE-CCSD data was re-plotted based on calculations from Ref.~\citenum{reinholdt2021nitrogen},
        and the experimental data was digitized from Ref.~\citenum{ekimova2017aqueous}
        using Web Plot Digitizer\cite{WebPlotDigitizer}.}
    \label{fig:Pople_only}
\end{figure*}
\begin{figure*}[htb]
    \centering
    \includegraphics[width=\textwidth]{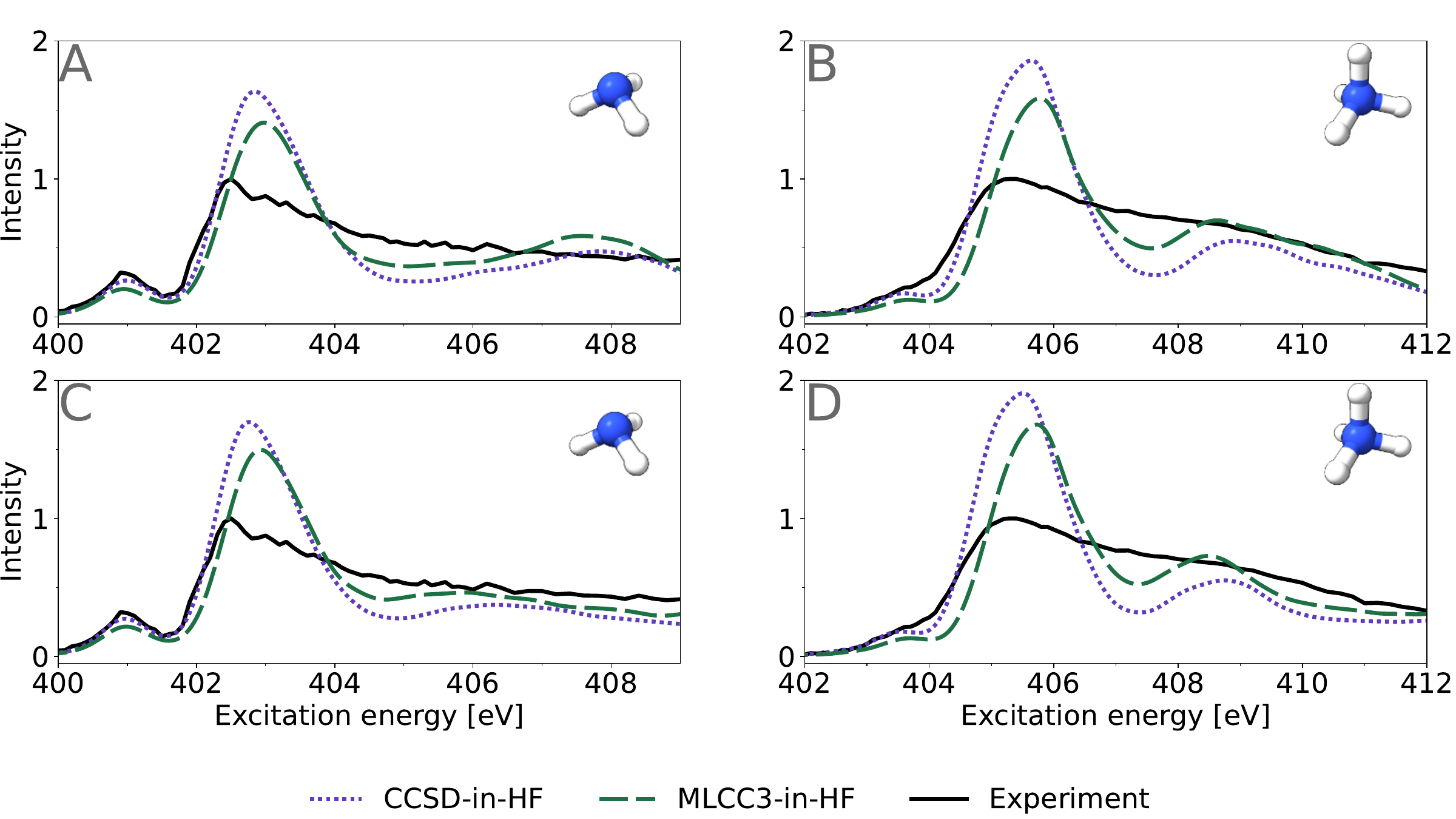}
    \caption{
        Comparison of XA spectra computed using Pople (Panels A, B)
        and Dunning (Panels C, D) basis sets.
        XA spectra of ammonia  (Panels A, C) and ammonium  (Panels C, D) in water clusters
        at the CCSD-in-HF and MLCC3-in-HF levels of theory.
        In the upper row 6-311++G$^{\ast\ast}$ was used in the CCSD space and 6-31G$^{\ast\ast}$ for the remaining solvent,
        while in the lower row aug-cc-pVTZ was used for the solute, aug-cc-pVDZ for the four closest water molecules
        and cc-pVDZ for the remaining solvent.
        Constant energy shifts of $\SI{-2.5}{\eV}$ and $\SI{-2.0}{\eV}$ were applied for CCSD-in-HF
        and $\SI{-1.5}{\eV}$, $\SI{-1.0}{\eV}$ for MLCC3-in-HF for the Pople and Dunning bases, respectively.
    }
    \label{fig:Pople_vs_Dunning}
\end{figure*}

To compare with the study of Reinholdt \textit{et al.}, \cite{reinholdt2021nitrogen}
we performed calculations using the same basis set (6-311++G**) in the coupled cluster region.
The remaining water molecules were described using either PDE or HF embedding with the 6-31G** basis set.
The resulting spectra for both ammonia and ammonium are presented in Figure\,\ref{fig:Pople_only}.
We use a constant broadening with Voigt profiles of $\SI{0.2}{\eV}$ Lorentzian
full width at half maximum (FWHM)
and Gaussian standard deviation (yielding a total FWHM of \SI{\sim0.59}{\eV})
to obtain smooth theoretical spectra.
The average spectra are normalized such that the area underneath
the curve equals that of the experiment in
the ranges $\SI{400}{\eV}\,$--$\,\SI{408}{\eV}$ and $\SI{402.5}{\eV}\,$--$\,\SI{411}{\eV}$
for NH$_3$ and NH$_4^+$, respectively.
The theoretical spectra for ammonia were shifted
to align approximately with the pre-edge of
the experiment (see Figure~\ref{fig:Pople_only}).
For ammonium the same shifts were applied.

Comparing the theoretical spectra of ammonia and ammonium,
shown in
Figure\,\ref{fig:Pople_only},  we observe that the
spectral shapes of the PE-CCSD spectra agree better
with experiment compared to both PDE-CCSD and CCSD-in-HF.
Since both the PDE and HF embedding schemes offer a higher level
of theoretical description of the solvent molecules,
the favourable agreement of PE-CCSD with experiment is likely due to cancellation of errors.
For instance,
a density-based embedding scheme includes the effects of Pauli-repulsion,
which is not the case for PE.
We note that
PDE-CCSD and CCSD-in-HF exhibit very similar spectral features.
Both methods overestimate
the ratio between the main-edge (at $\SI{402.8}{\eV}$) and the post-edge
($\SI{403}{\eV}$ to $\SI{409}{\eV}$)
of the ammonia spectrum (see  Panels A and C of Figure\,\ref{fig:Pople_only}),
compared to experiment and PE-CCSD.

The main-edges in PE-CCSD and PDE-CCSD exhibit a shoulder at approximately
$\SI{404.5}{\eV}$, a feature not observed for CC-in-HF.
It has been shown for a water dimer that
PDE is overestimating the repulsion energy compared to HF.\cite{olsen2015polarizable}
Therefore, in CC-in-HF,
the transitions in this energy region are probably
shifted to lower energies so that the shoulder merges with the main-edge.
With the selected normalization scheme,
the intensity of the post-edge region
is underestimated by PDE and HF-embedding,
except for a very broad feature at $\SI{\sim408}{\eV}$.
This broad band is shifted to lower energies for CCSD-in-HF compared to PDE-CCSD.
In contrast to PE-CCSD, we do not observe a spectral feature at the ionization limit
at approximately $\SI{407}{\eV}$.
Similar to the study by Reinholdt \textit{et al.},\cite{reinholdt2021nitrogen}
we do not treat the continuum differently than the other regions of the spectrum.
Therefore,
we conclude that this spectral feature is more likely an artifact of the PE description,
rather than the result of the discretized description of the continuum.

Inclusion of approximate triple excitations in MLCC3-in-HF compared to CCSD-in-HF
yields an improved intensity ratio between the main- and post-edge
in the ammonia spectrum,
resulting in a smoother transition between the two regions.
The absolute peak positions in MLCC3-in-HF are more accurate than in CCSD-in-HF.

For aqueous ammonium (see panels B and D of Figure\,\ref{fig:Pople_only}),
the important spectral characteristics --
the shoulder between $\SI{403}{\eV}$ and $\SI{404}{\eV}$, the main-edge at $\SI{405.7}{\eV}$,
and the post-edge between $\SI{407}{\eV}$ and $\SI{412}{\eV}$ -- are
reproduced with PDE-CCSD and CCSD-in-HF.
However, compared to PE-CCSD,
the transition from main- to post-edge is characterized by a significant drop in intensity.
This intensity decrease is more pronounced for PDE-CCSD,
because the broad post-edge feature is shifted to higher energies compared to CCSD-in-HF.

Similar to what we observed for ammonia,
the relative intensities of the main- and post-edge
are improved going from CCSD-in-HF to MLCC3-in-HF for ammonium:
in the MLCC3-in-HF spectrum, the main-edge intensity is reduced
and the post-edge intensity is increased.
Additionally,
we observe a red-shift of the post-edge feature from CCSD-in-HF to MLCC3-in-HF.
While the PE-CCSD spectrum agrees best with the experiment,
it contains a rather sharp feature in the range $\qtyrange{408}{410}{\eV}$ which is
much broader in the other embedding schemes.
This additional peak resembles the one in ammonia, and further supports
that it might be an artifact of the polarizable embedding.

To examine the basis set effect on the spectra,
we test various basis set combinations for the solute, the closest four waters
and the remaining solvent molecules
on an arbitrarily chosen snapshot.
The results are shown in Figures\,\ref{SI-fig:NH3_basis_sets}
and \ref{SI-fig:NH4_basis_sets} in the Supporting Information.
A combination of aug-cc-pVTZ on the solute,
aug-cc-pVDZ on the four closest water molecules,
and cc-pVDZ on the remaining solvent molecules
was chosen as a compromise between accuracy of the basis and computational efficiency.
Using these basis sets, the XA spectra of both ammonia and ammonium
were recomputed and compared with previous spectra (see Figure\,\ref{fig:Pople_vs_Dunning}).
For both ammonia and ammonium,
the change of basis set leaves the pre- and main-edges of the spectra unchanged.
However, for ammonia,
the post-edge feature is broadened
and red-shifted when the Dunning basis sets are used.
This change in spectral shape shows that the post-edge feature of ammonia
observed using the Pople basis
is not a real resonance, but rather an artifact of a limited basis set.
For ammonium, on the other hand,
the post-edge feature around $\SI{408.5}{\eV}$ does not shift,
and compression of the spectrum with increased basis set size is only observed above $\SI{409}{\eV}$.

\begin{figure}[htb]
    \centering
    \includegraphics[width=\linewidth]{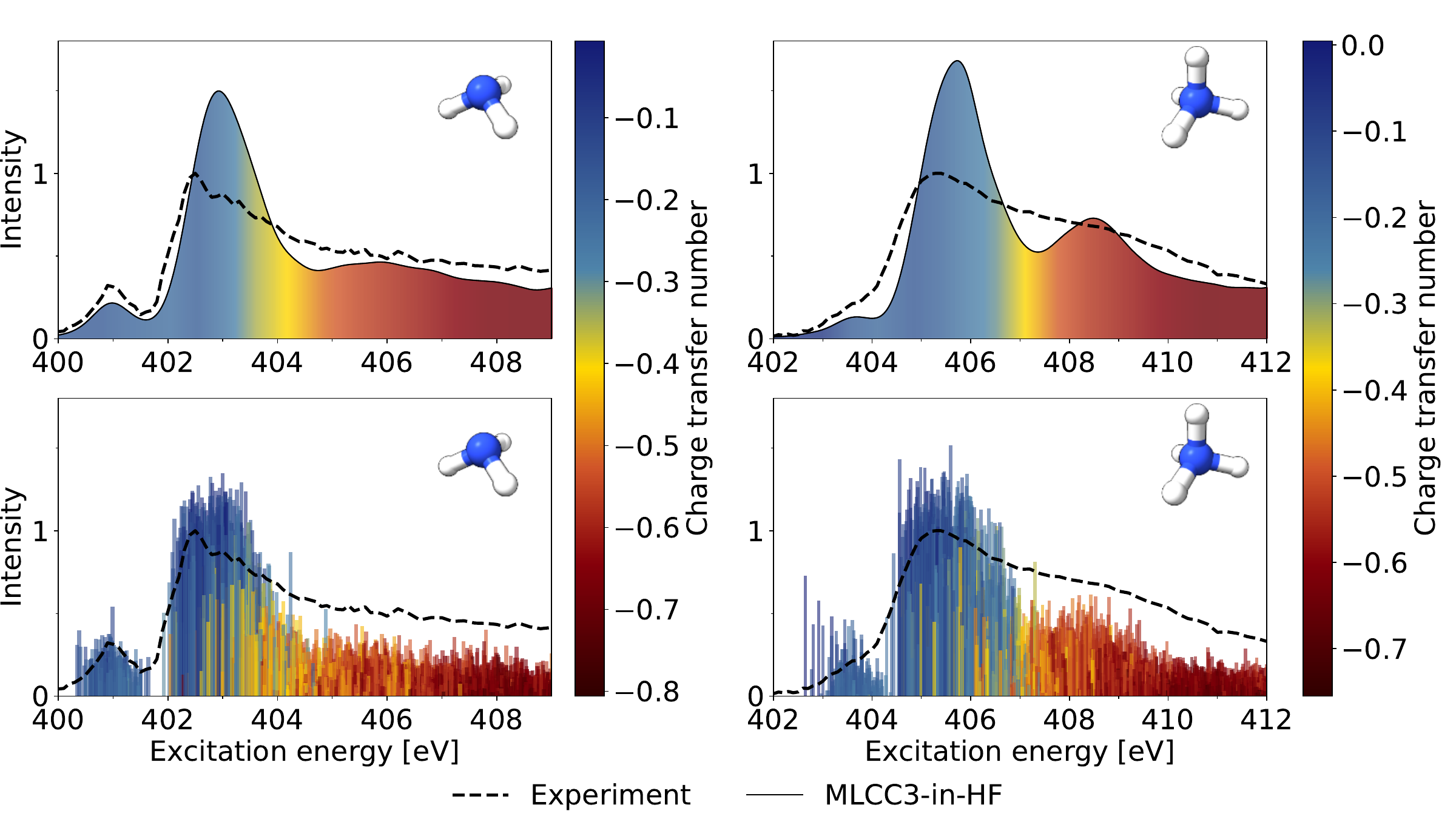}
    \caption{
        XA spectrum plotted together with the charge transfer number (CT) calculated
        for the solute molecules (ammonia, left column) and (ammonium right panel).
        In the lower panel, each excitation has been assigned a color based on its charge transfer character.
        In the upper panel, we show the broadened spectrum where the color gradient indicates the charge transfer character
        averaged for all transitions within a bin of the size equal to the FWHM. }
    \label{fig:CT_analysis}
\end{figure}

We have performed a charge transfer analysis for aqueous ammonia and ammonium solutions
with MLCC3-in-HF.
This analysis is based on the calculation of the so-called
charge transfer number for each excited state of the solute molecule.\cite{ChargeTransferNumbers0,ChargeTransferNumbers1,
ChargeTransferNumbers2,ChargeTransferNumbers3}
The charge transfer number
shows
the significance of the interaction with the solvent.
A negative value indicates the removal of electrons from the solute
while a positive value implies their addition.
Further details on the charge transfer analysis
can be found in Supporting Information and Ref.~\citenum{folkestad2023enhanced}.
The results are presented in Figure~\ref{fig:CT_analysis}.
As described above, and evident from the similarity in the respective values
for NH$_3$ and NH$_4^+$,
the charge transfer number acts as an indicator of environmental influence rather than actual electron transfer.
From Figure~\ref{fig:CT_analysis} we see that the excitations in
the pre- and main-edge in the XA spectra of both molecular systems are localized
on the respective solute molecules.
In the post-edge regions, we observe an increasing charge transfer indicating
larger influence from the solvent molecules.
This reflects the more diffuse nature of excitations in these regions.

In Figure\,\ref{SI-fig:density_analysis_AplusB} in the Supporting Information we show the charge transfer from the solute
and four surrounding water molecules to the remaining solvent.
We see that the excitations responsible for generating the pre-edge, main-edge,
and the first half of the post-edge are well described by the
four water molecules in the coupled cluster region.
An increasing degree of charge transfer in the second half of the post-edge is observed.
However, it is important to note
that the charge transfer analysis is only performed on the orbitals within the coupled cluster region.
Consequently, effects beyond the four nearest water molecules can only be included
through the diffuse basis functions and at the Hartree--Fock level.

To ensure the comparability with the results by Reinholdt \textit{et al.},\citep{reinholdt2021nitrogen}
we have included only four water molecules in the coupled cluster region.
However,
it is natural to increase the number of solvent molecules in the coupled cluster region
to examine the potential improvement in the spectra.
This is particularly interesting in the case of ammonium, where
the number of water molecules in the first solvation shell
is known to be 5.2.\cite{aydin2020similarities}
We have calculated the spectra of aqueous ammonium and ammonia with six water molecules in the coupled cluster region for a subset of 30 snapshots.
The results are presented in Figures\,\ref{SI-fig:NH3_4_vs_6_water} and \ref{SI-fig:NH4_4_vs_6_water} in the Supporting Information.
Although indicating that increasing the coupled cluster region could improve
the imbalance between the intensities of the main and post edges,
no significant changes in the spectra are observed.
These results are in agreement with the charge transfer analysis (see Figure\,\ref{SI-fig:density_analysis_AplusB}), indicating that
the CC region is sufficiently large.
However, a further improvement of the electronic structure,
achievable through a higher level of coupled cluster theory,
or improved quality of the structures
can further reduce the discrepancy between theory and experiment.
The quality of the structures can be, for example,
improved by accounting for nuclear quantum effects in the molecular dynamics simulations,
which play an important role in capturing the delicate nature of hydrogen bonds.\cite{ceriotti2016nuclear}
From a recent study on the XA of liquid water,\cite{folkestad2023enhanced}
we have observed both a redistribution of intensity from the main- to the post-edge
and a broadening effect from the inclusions of NQEs.

\section{Conclusions}
We have compared the performance of the different embedding schemes
---PE-CCSD, PDE-CCSD, CCSD-in-HF and MLCC3-in-HF---in simulating the XA spectra.
Our comparison reveals similarities
between PDE-CCSD and CCSD-in-HF, while PE-CCSD results differ noticeably.
Given that PDE-CCSD represents a theoretical refinement of PE-CCSD\cite{olsen2015polarizable, hrsak2018polarizable} and CC-in-HF methods are fully quantum mechanical,
we conclude that the close agreement of PE-CCSD with experiment is due to error cancellation.
Approaches that treat the environment through a density-type embedding will avoid issues relating to the lack of Pauli-repulsion,
as is a concern with PE.

A charge transfer analysis demonstrates the local character of the
excitations responsible for the pre- and main-edge,
in contrast to the more diffuse nature of post-edge excitations in both solutions.
Exploring the impact of an increased number of solvent molecules in the coupled cluster region
did not provide any significant improvement.
Therefore, we conclude that our methods offer a robust description of the
electronic structure of the solute,
with potential for improvement by using state-of-the-art path integral molecular dynamics trajectories or through higher-level electronic structure models.

\begin{acknowledgement}

This work has received funding from the European Research Council (ERC)
under the European Union’s Horizon 2020 Research and Innovation Program
(grant agreement No. 101020016).
S.D.F., A.C.P, and H.K. acknowledge funding from the
Research Council of Norway through FRINATEK (project No. 275506).
{M.O. acknowledges funding from the Swedish Research Council (grant agreement No. 2021-04521), and from the European Union's Horizon 2020 research and innovation programme under the Marie Sk{\l}odowska-Curie (grant agreement No. 860553).
S.C. acknowledges support from the Independent Research Fund Denmark--Natural Sciences (grant No. 7014-00258B).}
We acknowledge computing resources through UNINETT Sigma2
--the National Infrastructure for High Performance Computing
and Data Storage in Norway (project No. NN2962k).

\end{acknowledgement}

\begin{suppinfo}
The Supporting Information contains an examination of the basis set requirements on one arbitrarily selected snapshot,
charge transfer analysis on the solute and the four closest solvent molecules,
XA spectra obtained using various broadening schemes
(Gaussian and Lorentzian),
comparison of XA spectra obtained when four vs. six water molecules are included in the coupled cluster region.
Geometries are available from \url{https://doi.org/10.5281/zenodo.10390676}.

\setcounter{section}{0}%
\setcounter{figure}{0}%
\setcounter{table}{0}%
\setcounter{equation}{0}%
\renewcommand{\thesection}{S\arabic{section}}
\renewcommand{\thefigure}{S\arabic{figure}}
\renewcommand{\thetable}{S\arabic{table}}
\renewcommand{\theequation}{S\arabic{equation}}

\newpage

\section{Basis Set Study}

\begin{figure}[hbpt!]
    \centering    \includegraphics[height=0.8\textheight]{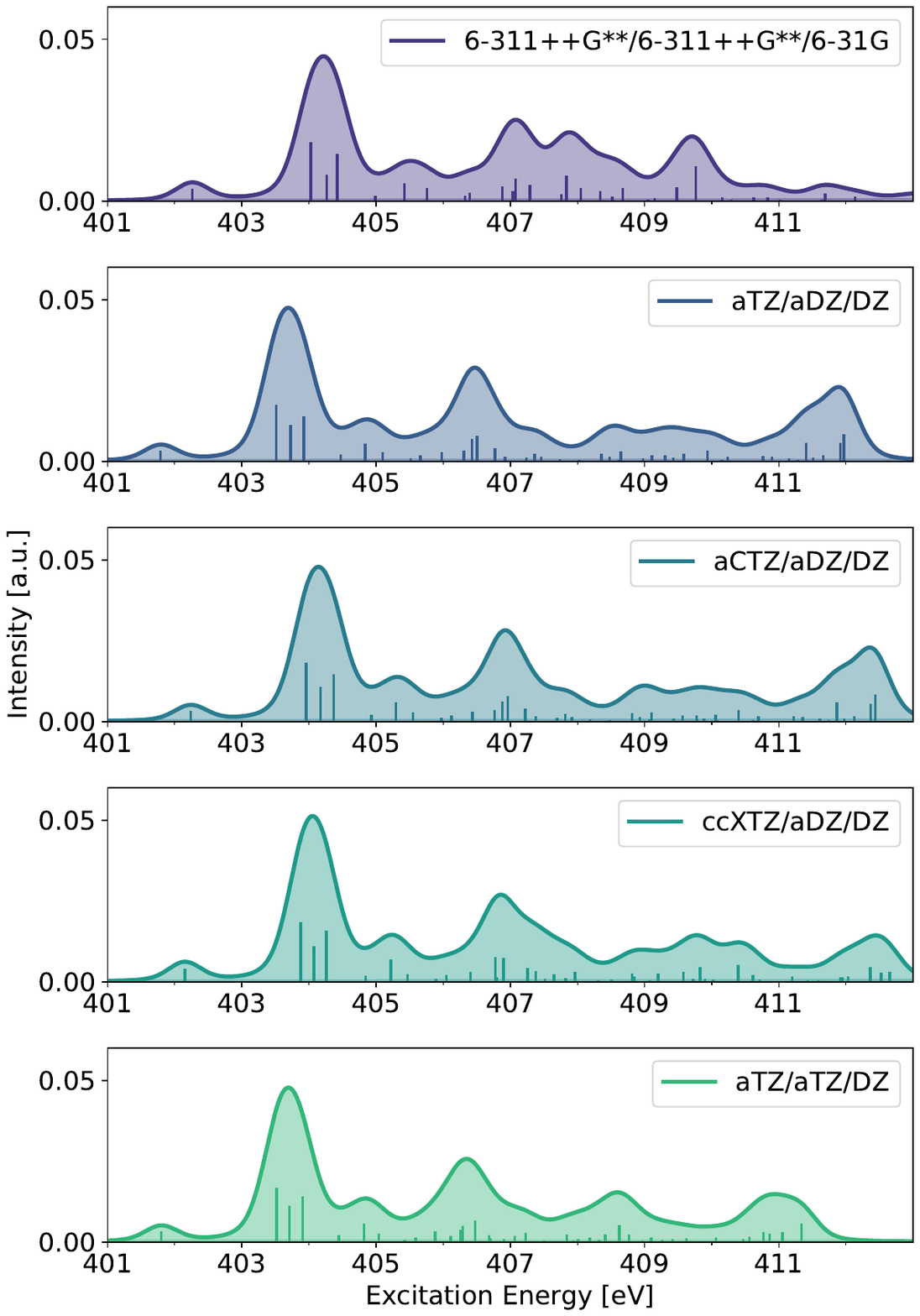}
    \caption{
        Comparison of different basis sets for the XA spectrum of \textbf{ammonia} in water at the MLCC3-12/72 level of theory
        using a single snapshot of the dynamics simulation (step 2200).
        Individual excitations have been broadened using Voigt profiles with $\SI{0.2}{\eV}$ Lorentzian fwhm
        and $\SI{0.2}{\eV}$ Gaussian standard deviation (total fwhm $\SI{\sim0.59}{\eV}$).
    }
    \label{SI-fig:NH3_basis_sets}
\end{figure}

\begin{figure}[hbpt!]
    \centering
\includegraphics[height=0.8\textheight]{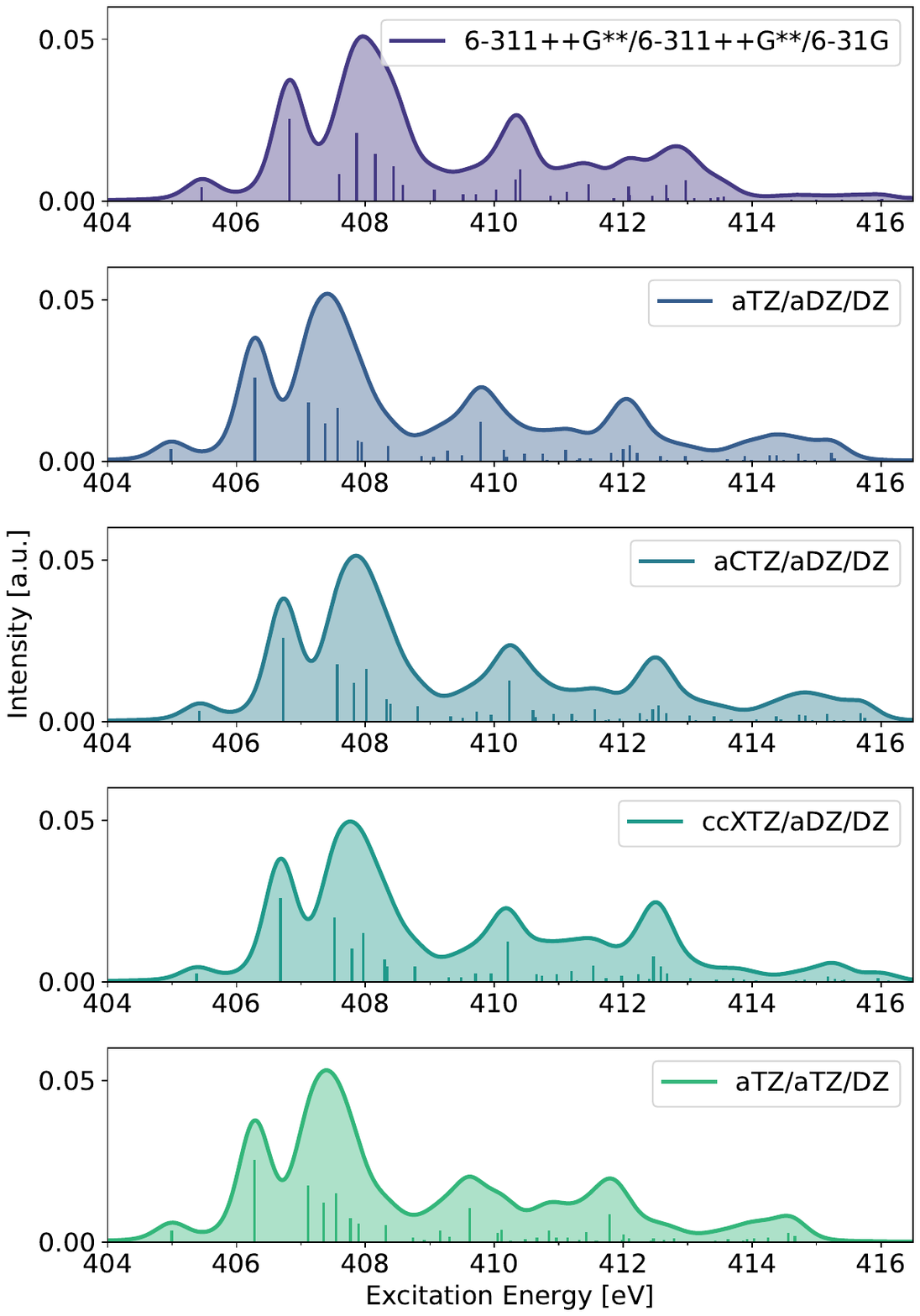}
    \caption{
        Comparison of different basis sets for the XA spectrum of \textbf{ammonium} in water at the MLCC3-12/72 level of theory
        using a single snapshot of the dynamics simulation (step 2200).
        Individual excitations have been broadened using Voigt profiles with $\SI{0.2}{\eV}$ Lorentzian fwhm
        and $\SI{0.2}{\eV}$ Gaussian standard deviation (total fwhm $\SI{\sim0.59}{\eV}$).
    }
    \label{SI-fig:NH4_basis_sets}
\end{figure}

\newpage

\mbox{}
\vspace{-0.1em}
\section{Charge Transfer Analysis}
\begin{figure}[hbpt!]
    \centering
    \includegraphics[width=\linewidth]{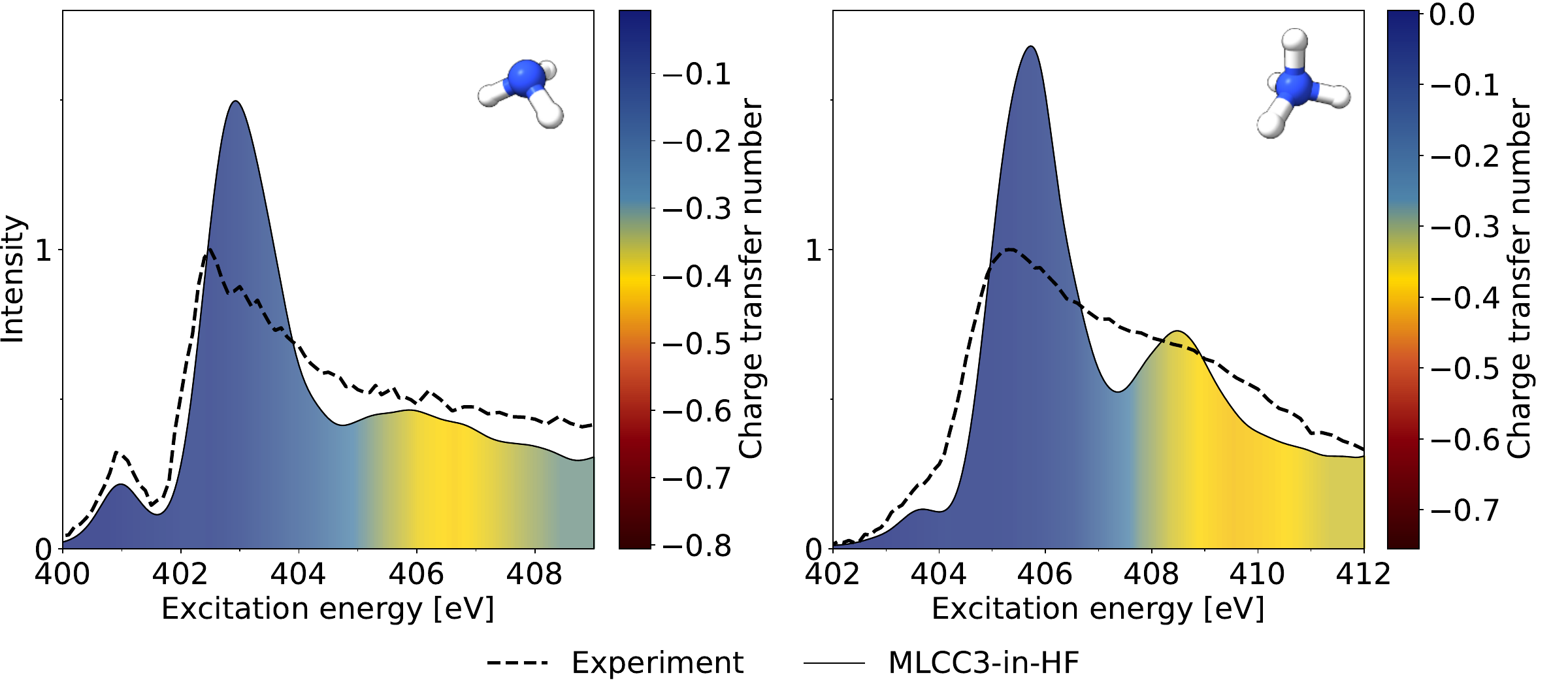}
    \caption{
        Spectrum plotted together with the charge transfer number for charge transfer from the central molecule and the 4 closest water molecules
        (ammonia (left) and ammonium (right)).
        The color map is identical to the one in Figure 3 of the main paper.
    }
    \label{SI-fig:density_analysis_AplusB}
\end{figure}

\newpage

\section{Increasing the number of solvent molecules in CC region}
\begin{figure}
    \centering
    \includegraphics[width=0.65\linewidth]{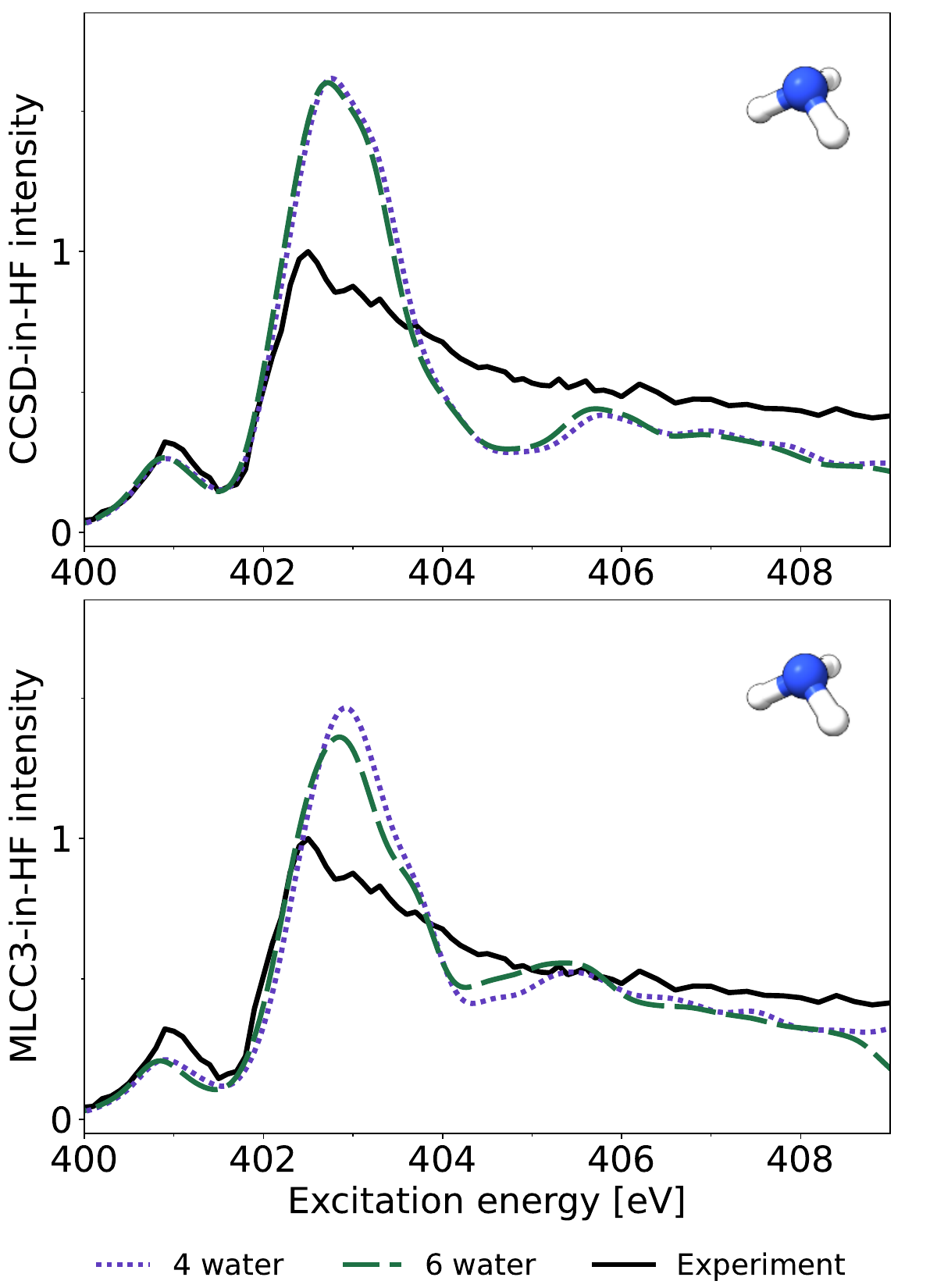}
    \caption{The effect of including 6 closest water molecules in CC-in-HF calculations for NH$_3$.
        Average over 30 snapshots.
        Voigt profiles with Lorentzian fwhm of 0.2~eV and 0.2 Gaussian standard deviation.
        XA spectra of \textbf{ammonia} in water clusters at the CCSD-in-HF and MLCC3-in-HF level of theory.
        The spectra have been shifted by $\SI{-2}{\eV}$ and $\SI{-1}{\eV}$, respectively.
    }
    \label{SI-fig:NH3_4_vs_6_water}
\end{figure}

\begin{figure}
    \includegraphics[width=0.65\linewidth]{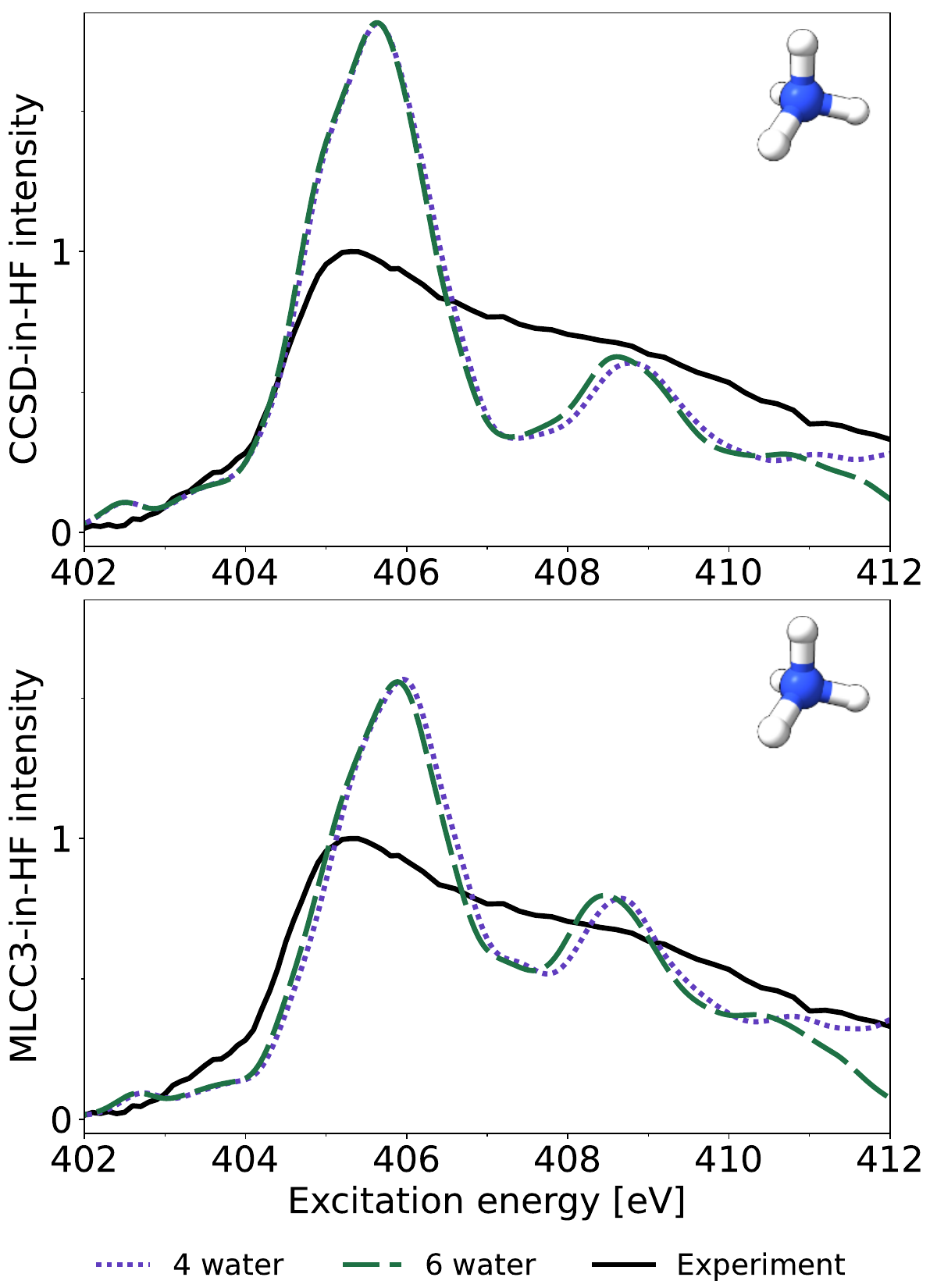}
    \caption{
        The effect of including 6 closest water molecules in CC-in-HF calculations for NH$_4^+$.
        Average over 30 snapshots.
        Voigt profiles with Lorentzian fwhm of 0.2~eV and 0.2 Gaussian standard deviation.
        XA spectra of \textbf{ammonium} in water clusters at the CCSD-in-HF and MLCC3-in-HF level of theory.
        The spectra have been shifted by $\SI{-2}{\eV}$ and $\SI{-1}{\eV}$, respectively.
        }
    \label{SI-fig:NH4_4_vs_6_water}
\end{figure}

\newpage
\section{Convergence from CCSD to CC3}

\begin{figure}[hbpt!]
    \centering
\includegraphics[height=0.8\textheight]{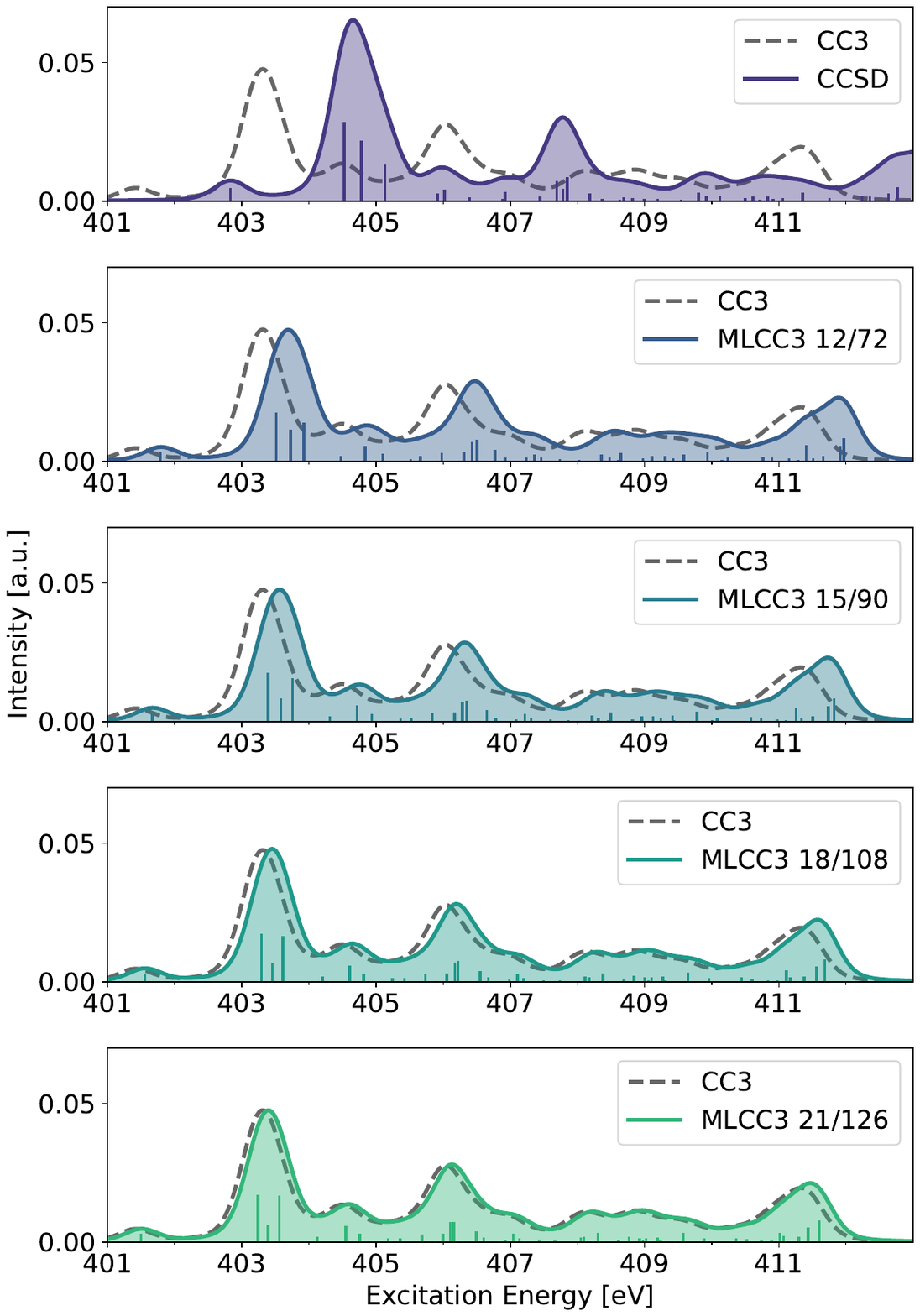}
    \caption{
        Convergence from CCSD to CC3 by increasing the number of occupied/virtual CNTOs included in the MLCC3 orbital space.
        The spectra are obtained for a single snapshot of the dynamics simulation of \textbf{ammonia} (step 2200).
        Individual excitations have been broadened using Voigt profiles with $\SI{0.2}{\eV}$ Lorentzian fwhm
        and $\SI{0.2}{\eV}$ Gaussian standard deviation (total fwhm $\SI{\sim0.59}{\eV}$).
    }
    \label{fig:NH3_mlcc3_space}
\end{figure}

\begin{figure}
    \centering
    \includegraphics[height=0.85\textheight]{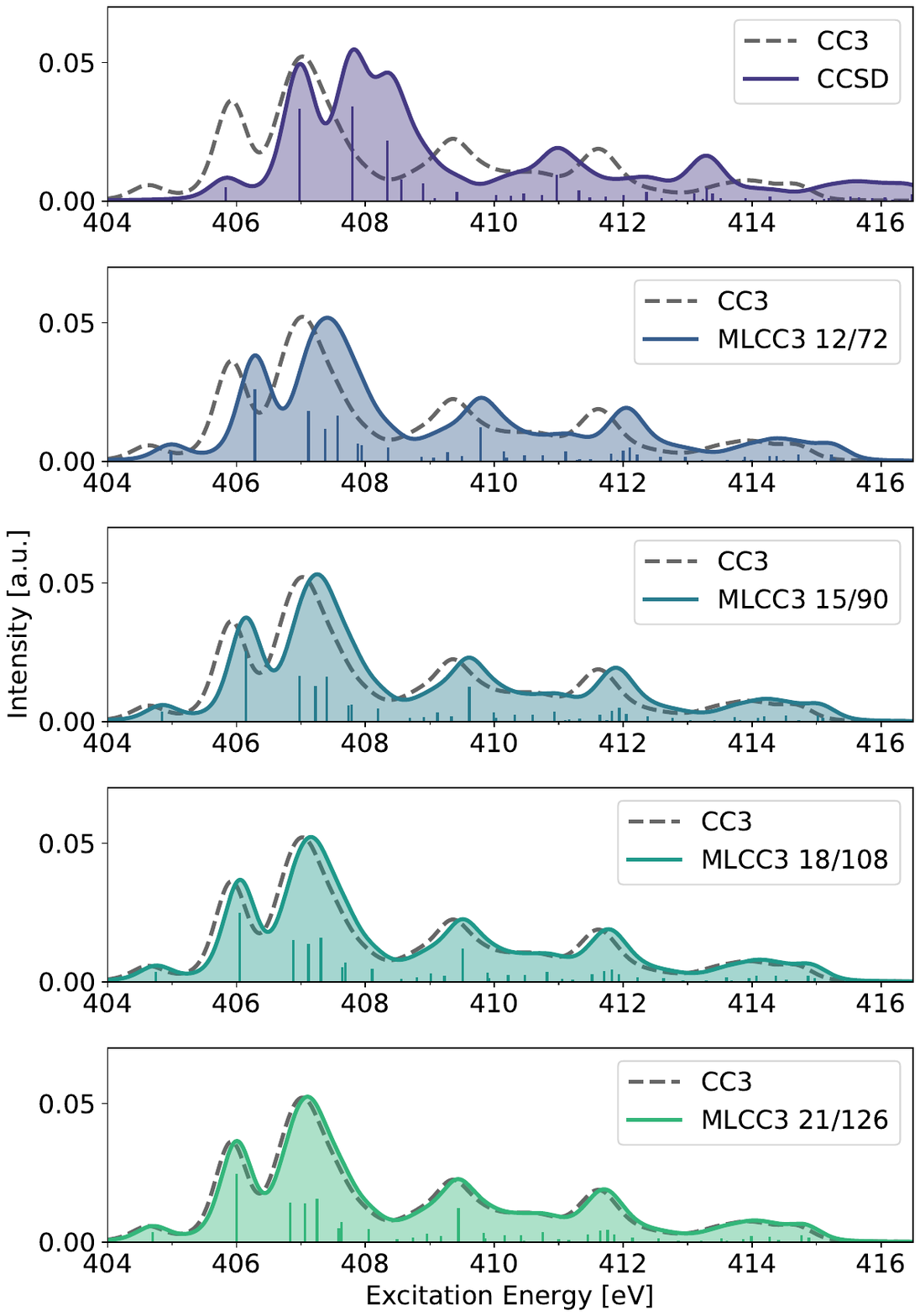}
    \caption{
        Convergence from CCSD to CC3 by increasing the number of occupied/virtual CNTOs included in the MLCC3 orbital space.
        The spectra are obtained for a single snapshot of the dynamics simulation of \textbf{ammonium} (step 2200).
        Individual excitations have been broadened using Voigt profiles with $\SI{0.2}{\eV}$ Lorentzian fwhm
        and $\SI{0.2}{\eV}$ Gaussian standard deviation (total fwhm $\SI{\sim0.59}{\eV}$).
    }
    \label{fig:NH4_mlcc3_space}
\end{figure}

\newpage

\section{XA spectra using various broadening schemes}

\begin{figure}[hbtp!]
    \centering
    \includegraphics[width=0.8\textwidth]{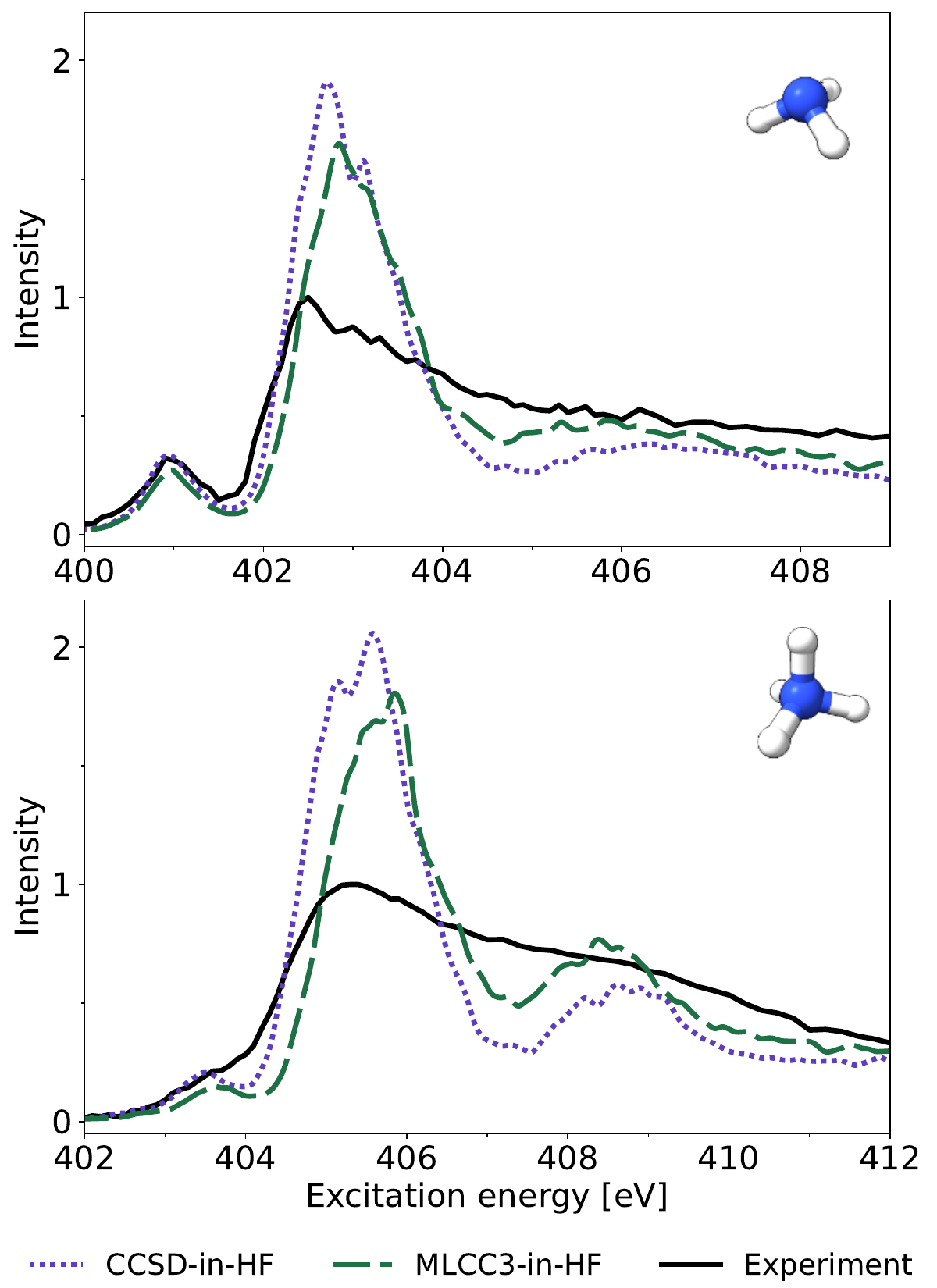}
    \caption{
        XA spectra of NH$_3$ (top) and NH$^+_4$ (bottom) in water clusters
        at the CCSD-in-HF and MLCC3-in-HF levels of theory.
        aug-cc-pVTZ was used for the solute,
        aug-cc-pVDZ for the four closest water molecules
        and cc-pVDZ for the remaining solvent.
        Broadening: Lorentzian profiles with $\SI{0.2}{\eV}$ fwhm.
        Constant energy shifts of $\SI{-2.0}{\eV}$ and $\SI{-1.1}{\eV}$
        were applied for CCSD-in-HF and MLCC3-in-HF respectively.
    }
    \label{fig:SI-Lorentzian}
\end{figure}

\begin{figure}
    \centering
\includegraphics[width=0.8\textwidth]{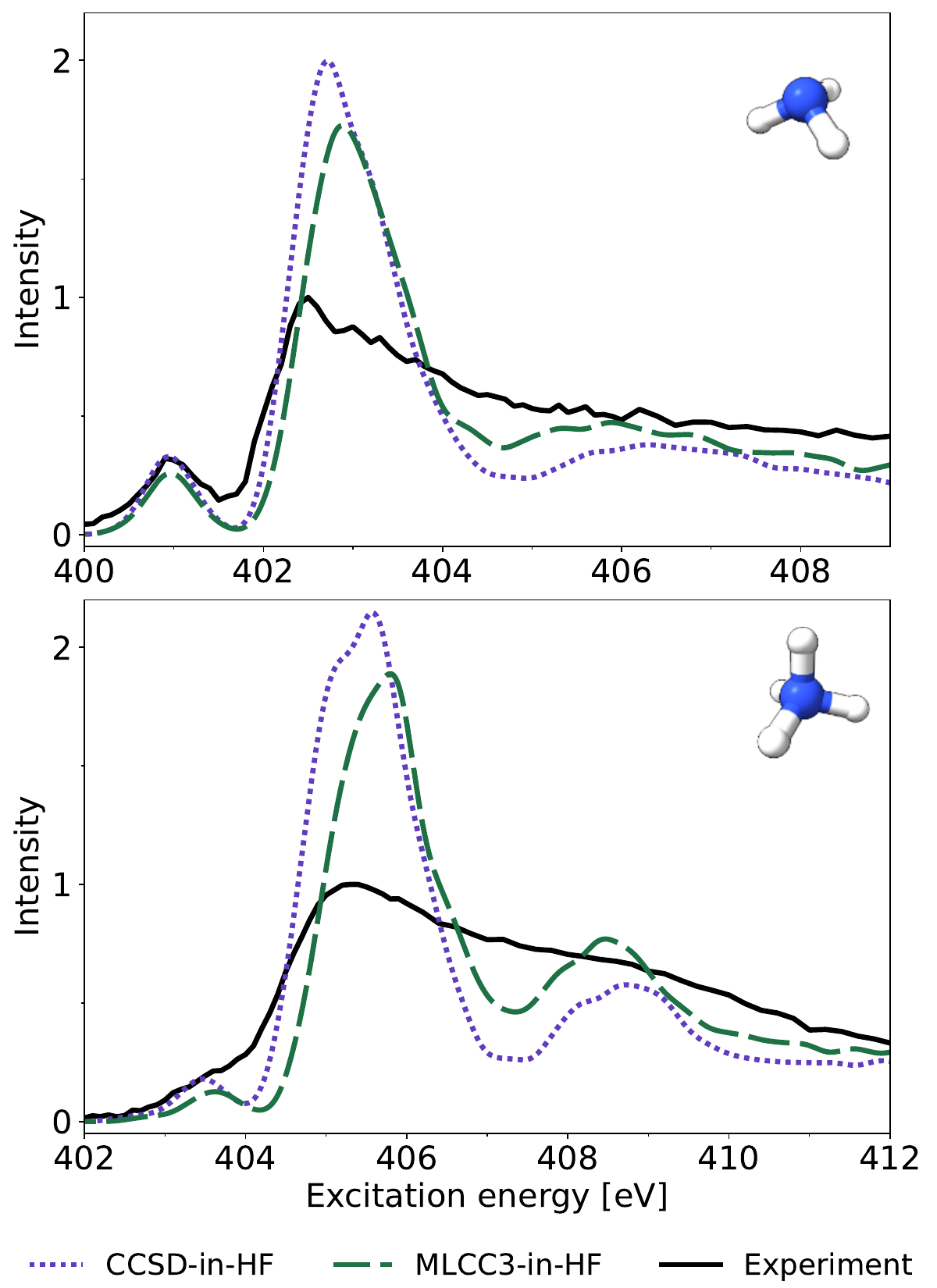}
    \caption{
        XA spectra of NH$_3$ (top) and NH$^+_4$ (bottom) in water clusters
        at the CCSD-in-HF and MLCC3-in-HF levels of theory.
        aug-cc-pVTZ was used for the solute,
        aug-cc-pVDZ for the four closest water molecules
        and cc-pVDZ for the remaining solvent.
        Broadening: Gaussian profiles with $\SI{0.4}{\eV}$ fwhm.
        Constant energy shifts of $\SI{-2.0}{\eV}$ and $\SI{-1.1}{\eV}$
        were applied for CCSD-in-HF and MLCC3-in-HF respectively.
    }
    \label{fig:SI-Gaussian}
\end{figure}

\end{suppinfo}

\bibliography{paper}

\end{document}